\documentclass[12pt]{article}
\usepackage{graphicx}
\usepackage{amsfonts}
\usepackage{amssymb}
\usepackage{amsmath}
\usepackage{latexsym}
\usepackage{color}
\usepackage{cite}
\input{colordvi.tex}

\renewcommand{\thefootnote}{\#\arabic{footnote}}

\newcommand{\bea}{\begin{eqnarray}}  \newcommand{\eea}{\end{eqnarray}}
\newcommand{\beq}{\begin{equation}}  \newcommand{\eeq}{\end{equation}}

\begin{document}

\setcounter{footnote}{0}
\begin{titlepage}

\begin{center}

\vskip .5in

{\Large \bf Primordial Non-Gaussianity in Models with

Dark Matter Isocurvature Fluctuations}

\vskip .45in

{\large
Tomo Takahashi$^1$, Masahide Yamaguchi$^{2}$
and Shuichiro Yokoyama$^3$
}

\vskip .45in

{\em
$^1$
Department of Physics, Saga University, Saga 840-8502, Japan \\
$^2$Department of Physics and Mathematics, Aoyama Gakuin
University, Sagamihara 229-8558, Japan \\
$^3$Department of Physics and Astrophysics, Nagoya University,
Aichi 464-8602, Japan
}

\end{center}

\vskip .4in

\begin{abstract}
  We investigate primordial non-Gaussianity and dark matter isocurvature
  fluctuations in the modulated reheating and the curvaton scenarios.  In
  these scenarios, large non-Gaussianity can be generated, on the other
  hand, depending on how dark matter is produced, too large isocurvature
  fluctuations can also arise, which is  inconsistent with current
  observations.  In this paper, we study this issue in a mixed scenario
  where the curvature fluctuations can also be produced from the
  inflaton fluctuations as well as those from a light scalar field such
  as the modulus and the curvaton. We show that 
  primordial fluctuations can be highly non-Gaussian 
   without conflicting the current constraint on
  isocurvature fluctuations for such mixed scenarios. However, if the
  constraint on isocurvature fluctuations becomes
  severer as expected by the Planck satellite, $f_{\rm NL}$, 
  a nonlinearity parameter for adiabatic fluctuations, 
  should be very small as $f_{\rm NL} \lesssim 3$, which would give 
  interesting implications for the generation mechanism of 
  dark matter. Non-Gaussianity from isocurvature fluctuations 
  is also discussed in these scenarios.
\end{abstract}

\end{titlepage}

\renewcommand{\thepage}{\arabic{page}}
\setcounter{page}{1}
\renewcommand{\thefootnote}{\#\arabic{footnote}}

%%%%%%%%%%%%%%%%%%%%%%%%%%%%%%%%%%%%%%%%%%%%%%
\section{Introduction}\label{sec:int}
%%%%%%%%%%%%%%%%%%%%%%%%%%%%%%%%%%%%%%%%%%%%%%

Cosmic density fluctuations that we can observe today originate to those
generated at the early universe.  Since current cosmological
observations on density fluctuations are so precise, they provide us a
lot of information on the physics of the early universe.  Among various
observables, non-Gaussianity of primordial perturbations has been
attracting much attention recently. One of the reason is that current
and upcoming cosmological observations can well probe the Gaussian
nature of perturbations more accurately than before. The standard
simple inflation model predicts almost Gaussian primordial fluctuations
and its deviation from the Gaussian fluctuations is less than
$10^{-5}$. Thus, larger deviation from the Gaussian fluctuations, which
is still less than $0.1\%$ of the Gaussian part, would indicate that we
need some mechanism of generating primordial curvature
perturbations other than the standard single slow-roll field inflation. The size of non-Gaussianity is
usually characterized by a non-linearity parameter $f_{\rm NL}$.  The
purely Gaussian fluctuations correspond to $f_{\rm NL}=0$ and the
current constraint on this quantity is given as $-9 < f_{\rm NL} < 111$
at $95\%$ C.L. in Ref.~\cite{Komatsu:2008hk} and $-4 < f_{\rm NL} < 80$
at $95\%$ C.L. in Ref.~\cite{Smith:2009jr}\footnote{
  Two types of non-Gaussianity are often discussed in the literatures,
  one is the so-called local type and the other is the equilateral
  type.  In this paper, we only consider the local type non-Gaussianity.
}.

Another test of primordial fluctuations is the adiabaticity of
primordial fluctuations.  If any deviation from purely adiabatic
fluctuations is discovered, it would have important implications for
the generation mechanism of matter (dark matter (DM) and baryon) in the early
universe.  The adiabatic relation between matter and radiation should
be satisfied when both of them are created from
a single component. If, however, the deviation
from the adiabatic relation is detected, then it implies that
radiation and matter have originated from separate components. To
quantify the deviation from the adiabaticity, the fraction of the
isocurvature fluctuations to the total ones is usually used,
which is defined with their power spectra at some reference scale and
denoted as $\alpha$. Depending on how isocurvature fluctuations are generated,
 such an
isocurvature mode can be correlated/uncorrelated with the adiabatic
ones. 
The current observational limits on $\alpha_0$ (uncorrelated type) 
and $\alpha_{-1}$ (correlated type) are given by
$\alpha_0 < 0.16~ (0.072) $ at $95\%$ C.L. from WMAP5-only
(from WMAP5+BAO+SN) and
$\alpha_{-1} < 0.011~(0.0041)$ at $95\%$ C.L. from WMAP5-only
(from WMAP5+BAO+SN), respectively \cite{Komatsu:2008hk}.

As a possible mechanism of generating large non-Gaussianity, the
curvaton \cite{Mollerach:1989hu,Enqvist:2001zp} and the modulated
reheating scenarios \cite{Dvali:2003em} have been investigated in various
contexts. Although these mechanisms are attractive with regard to
producing large non-Gaussianity, when one considers the generation of baryon asymmetry
and DM in these mechanisms, large isocurvature fluctuations
may arise \cite{Moroi:2002rd,Lyth:2002my,Lyth:2003ip,Beltran:2008ei,Moroi:2008nn,Takahashi:2009dr}, which would indicate that such scenarios are
disfavored by cosmological observations.

In our previous letter \cite{Takahashi:2009dr}, we have investigated
density fluctuations in a scenario with gravitino DM in the
framework of modulated reheating \cite{Dvali:2003em}, 
which is known to generate
large non-Gaussianity \cite{Zaldarriaga:2003my}. Then, we
have shown that gravitino DM is disfavored if the adiabatic
curvature perturbations have large local-type non-Gaussianity because of
generating too large DM isocurvature fluctuations
simultaneously\footnote{
In fact, in the framework of the curvaton scenario \cite{Mollerach:1989hu,Enqvist:2001zp}, which is also known as  a good candidate to produce large non-Gaussianity
\cite{nonG_curvaton}, 
such DM scenario may  not be viable because of too large isocurvature 
fluctuations as well.
}. 
However, in general, fluctuations from the inflaton can also contribute to 
cosmic density fluctuations today even if we consider mechanisms 
such as the modulated reheating and 
the curvaton. In such a case, density fluctuations are 
a mixture of fluctuations originating from multiple sources.
Such a mixed scenario in the framework of the
modulated reheating \cite{Ichikawa:2008ne} and the curvaton
\cite{Lazarides:2004we,Langlois:2004nn,Moroi:2005kz,Moroi:2005np,Ichikawa:2008iq,Langlois:2008vk} has been discussed by several authors. With the
contribution from the inflaton fluctuations, it is expected that the
situation could be dramatically changed because both the isocurvature
fluctuations and the non-Gaussianity are suppressed. In this paper, we
consider the density fluctuations in a scenario with dark
matter isocurvature fluctuations in the mixed modulated reheating and 
the mixed curvaton scenarios.
Then, we investigate whether large non-Gaussianity can be
generated without conflicting  observational limits on
isocurvature fluctuations. Although 
we focus on the gravitino (axino) DM scenario when we discuss 
the issue in the framework of the mixed modulated reheating,
our discussion on the mixed curvaton scenario 
can apply to a generic DM which originates from the inflaton or the
curvaton.  We also discuss non-Gaussianity from isocurvature fluctuations in these scenarios, which
can also affect the non-linearity of cosmic microwave background (CMB) temperature fluctuations.

The paper is organized as follows. In section~\ref{sec:form}, we
summarize a formulation to investigate non-Gaussianity based on $\delta
N$ formalism and  cold dark matter (CDM) isocurvature
fluctuations. In section~\ref{sec:mod}, we briefly review the modulated
reheating scenario.  In section~\ref{sec:mixmod}, we investigate
non-Gaussianity and the gravitino DM isocurvature fluctuations
in the mixed modulated reheating scenario. In section~\ref{sec:curv}, we
move on to the discussion on the mixed curvaton scenario.  In
section~\ref{sec:iso}, we discuss the non-linearity of isocurvature
fluctuations in these scenarios. Section~\ref{sec:sum} summarizes our
results. Throughout this paper, we set the reduced Planck mass $M_{\rm
Pl}^2 = (8\pi G)^{-1}$ to be unity, where $G$ is the gravitational
constant.

%%%%%%%%%%%%%%%%%%%%%%%%%%%%%%%%%%%%%%%%
\section{Formulation}\label{sec:form}
%%%%%%%%%%%%%%%%%%%%%%%%%%%%%%%%%%%%%%%%

In this section, we summarize a formalism to discuss primordial
non-Gaussianity and isocurvature fluctuations.

%%%%%%%%%%%%
\subsection{Non-linearity parameter in $\delta N$ formalism}\label{subsec:NG}
%%%%%%%%%%%%

First, we give the definition of a non-linearity parameter which
characterizes non-Gaussianity of the primordial curvature
perturbations.  In order to evaluate the curvature perturbations
$\zeta$ on super-horizon scales, we adopt the $\delta N$ formalism
\cite{deltaN}. In this formalism, the curvature
perturbations on sufficiently large scales $\zeta$ at the final time
$t=t_f$ are identical to the perturbations of the $e$-folding number
measured in the homogeneous FRW Universe from the initial time $t=t_*$
to the final time $t_f$ as
\begin{eqnarray}
\zeta (t_f) \simeq \delta N (t_f,t_*)~,
\end{eqnarray}
where $N$ represents the $e$-folding number defined as $N(t_f,t_*) =
\int^{t_f}_{t_*}Hdt$ with $H$ being the Hubble parameter.  Usually a
final hypersurface at $t=t_f$ is taken to be a uniform energy density
one and an initial hypersurface at $t=t_*$ to be a flat one.  Taking
the initial time to be some time shortly after horizon crossing
during inflation, we can expand $\delta N$ in terms of
fluctuations of scalar fields $\varphi^a$ on the initial flat
hypersurface as
\begin{eqnarray}
\zeta (t_f) \simeq N_a \delta \varphi^a_* + {1 \over 2}N_{ab}\delta
 \varphi^a_* \delta \varphi^b_*~,
\label{eq:deltaN}
\end{eqnarray}
up to the second order. Here, a superscript $a$ labels a scalar field
and $N_a \equiv \partial N(t_f,t) / \partial \varphi^a(t) |_{t=t_*}$
and $N_{ab} \equiv \partial^2 N(t_f,t) /\partial \varphi^a(t) \partial
\varphi^b(t) |_{t=t_*}$. The summation is implied for the repeated
indices.

To discuss non-Gaussianity, one usually considers the bispectrum
(3-point correlation function) of the curvature perturbations which is
written as
\begin{equation}
%\label{ }
\langle \zeta_{\vec{k}_1} \zeta_{\vec{k}_2} \zeta_{\vec{k}_3}\rangle
= (2\pi)^3 \delta^{(3)} \left( \vec{k}_1 + \vec{k}_2 + \vec{k}_3\right) 
B_\zeta (\vec{k}_1, \vec{k}_2, \vec{k}_3 ).
\end{equation}
To quantify the size of non-Gaussianity, the
non-linearity parameter $f_{\rm NL}$ is often adopted and is defined as
\begin{equation}
%\label{ }
B_\zeta (\vec{k}_1, \vec{k}_2, \vec{k}_3 ) = 
\frac{6}{5} f_{\rm NL} 
\left( P_\zeta(k_1)P_\zeta(k_2) + 2~{\rm perms.}\right).
\end{equation}
In the $\delta N$ formalism, we can write the 3 point function of the
curvature perturbation as
\begin{eqnarray}
\label{eq:3point}
\langle \zeta_{\vec{k}_1} \zeta_{\vec{k}_2} \zeta_{\vec{k}_3}\rangle
&\simeq & (2\pi)^3
\left[ N^a N^b N_{ab}  + 
N_{ab}N^{bc}N^a_{~c} {\cal P}_\delta \ln (k_m L) \right] 
\nonumber\\
&& \times
\left( P_\delta(k_1)P_\delta(k_2) + 2~{\rm perms.}\right) 
\delta^{(3)} \left( \vec{k}_1 + \vec{k}_2 + \vec{k}_3\right)~, 
\end{eqnarray}
where we have neglected the nonlinearity of $\delta \varphi^a_*$.
Here the indices are lowered and raised by using the Kronecker's delta
$\delta^{ab}$.  The second term in the parenthesis is a contribution
from the one-loop correction. $k_m \equiv {\rm min}\{k_i\}(i=1,2,3)$ and
$L$ is a cutoff scale which is often taken to be the order of the
present Hubble scale. $P_\delta$ and $P_\zeta$ represent power spectra of fluctuations of scalar
fields and the curvature perturbations, respectively. They are given 
and related to ${\cal P}_\delta$ and ${\cal P}_\zeta$ by 
\begin{eqnarray}
\label{eq:powerdelta}
&&
\langle \delta \varphi^a_{*\vec{k}_1}\delta \varphi^b_{*\vec{k}_2}\rangle
\equiv (2\pi)^3 \delta^{ab} \delta^{(3)}\left( \vec{k}_1 + \vec{k}_2\right)P_\delta(k_1)
= {2\pi^2 \over k_1^3}\delta^{ab}{\cal P}_\delta \delta^{(3)} \left( \vec{k}_1 + \vec{k}_2\right)~,\\
\label{eq:power}
&&\langle \zeta_{\vec{k}_1}\zeta_{\vec{k}_2}\rangle
\equiv (2\pi)^3  \delta^{(3)}\left( \vec{k}_1 + \vec{k}_2\right) P_\zeta(k_1)
= {2\pi^2 \over k_1^3}{\cal P}_\zeta \delta^{(3)} \left( \vec{k}_1 + \vec{k}_2\right)
~.
\end{eqnarray}
The relation between $P_\zeta$ and $P_\delta$ can be written as
\begin{eqnarray}
P_\zeta(k) = \left[ N_a N^a + N_{ab}N^{ab}{\cal P}_\delta \ln (k L) \right] P_\delta (k)~.
\end{eqnarray}
Then we can express the non-linearity parameter $f_{\rm NL}$ as
\begin{eqnarray}
{6 \over 5}f_{\rm NL}
= {1 \over \left( N_c N^c  + N_{cd}N^{cd}{\cal P}_\delta \ln (k_m L) \right)^2} 
\left[ N^a N^b N_{ab}
+ N_{ab}N^{bc}N^a_{~c}{\cal P}_\delta \ln (k_m L) \right]~.
\label{NLP}
\end{eqnarray}
Here, we neglect the contribution coming from the higher order in
$\delta \varphi^a_*$, for example,  terms with $N_{abc}$,
$N_{abcd}$, and so on.  We will justify this assumption for the
modulated reheating scenario later\footnote{
For the curvaton scenario, 
as shown in Ref.~\cite{SVW2006}, one can find that
the third order terms with $N_{abc}$ can be neglected
for large $f_{\rm NL}$
in the absence of the non-linear evolution of the curvaton
field between the horizon crossing and the start of curvaton oscillation.}. 

%%%%%%%%%%%%%%
\subsection{Cold dark matter (CDM) isocurvature fluctuations}\label{sec:cdm}
%%%%%%%%%%%%%%

Now let us move on to the issue of CDM isocurvature fluctuations.  If we
consider the curvature perturbations $\zeta_i$ on the spatial slices of
uniform density $\rho_i$ for the $i$-th component, which are related to
the total curvature perturbation $\zeta$ as $\zeta = \sum_i \zeta_{i}
\dot{\rho}_{i} / \dot{\rho}$, isocurvature fluctuations between CDM and
radiation are defined as
\begin{eqnarray}
S_{\rm CDM} \equiv 3 (\zeta_{\rm CDM} - \zeta_{\rm r}),
\label{iso}
\end{eqnarray}
where $\zeta_{\rm CDM}$ and $\zeta_r$ are the curvature perturbations
defined on the slice of $\rho_{\rm CDM}$ and $\rho_{r}$ being uniform,
respectively.  One can also write down this quantity using the
fluctuations of the ratio between the number density of CDM, $n_{\rm
CDM}$, and the entropy, $s$, as
\begin{eqnarray}
S_{\rm CDM} = 
{\delta (n_{\rm CDM} / s) \over n_{\rm CDM} / s}
= {\delta n_{\rm CDM} \over n_{\rm CDM}} - {\delta s \over s}.
\label{iso2}
\end{eqnarray}

To parametrize the contribution from isocurvature fluctuations, one
usually uses the fraction of isocurvature fluctuations to the total
ones which is defined as
\begin{equation}
\label{eq:alpha}
\alpha \equiv \frac{P_S ( k_0) }{P_\zeta (k_0) + P_S (k_0)},
\end{equation}
where 
$k_0$ is some reference scale at which the power spectra are evaluated.
$P_S(k)$ is the power spectrum for isocurvature fluctuations
defined by
\begin{equation}
%\label{eq:power_iso}
\langle S_{\vec{k}_1} S_{\vec{k}_2}\rangle
\equiv  (2\pi)^3 \delta^{(3)}\left( \vec{k}_1 + \vec{k}_2\right)P_S(k_1)
= {2\pi^2 \over k_1^3}{\cal P}_S \delta^{(3)} \left( \vec{k}_1 + \vec{k}_2\right).
\end{equation}
In the scenario discussed in the following, isocurvature fluctuations
can be correlated/uncorrelated with adiabatic ones and both can arise
simultaneously.  In such a case, we need to define the fraction $\alpha$
separately for correlated and uncorrelated ones, respectively. Detailed
discussion on this point will be made in the next section.

As in the case for the curvature (adiabatic) fluctuations, we can also
define the non-linearity parameters for isocurvature fluctuations
$f_{\rm NL}^{\rm (iso)}$ as
\begin{equation}
%\label{ }
S_{\rm CDM} = S_g + f_{\rm NL}^{\rm (iso)} S_g^2,
\end{equation}
where $S_g$ is the Gaussian part of isocurvature fluctuations.  In
some cases, the contribution from the second order term dominates over
the first order one. In such a case, the fraction of isocurvature
fluctuations $\alpha$ represents the size of
non-Gaussianity. Regarding the notation for non-Gaussianity from
isocurvature fluctuations, we follow those of
Ref.~\cite{Hikage:2008sk}.

%%%%%%%%%%%%%%%%%%%%%%%%%%%%%%%%%%%%%%%%
\section{Modulated reheating scenario}\label{sec:mod}
%%%%%%%%%%%%%%%%%%%%%%%%%%%%%%%%%%%%%%%%

In this section we give a brief review of the modulated reheating
scenario and some results for non-Gaussianity of the curvature
perturbations and gravitino DM isocurvature fluctuations in
this scenario.

%%%%%%%%%
\subsection{Non-Gaussianity of the curvature perturbations}\label{sec:NGmod}
%%%%%%%%%

Let us start with considering the background dynamics during reheating
era, in order to evaluate the curvature perturbations in the modulated
reheating scenario based on $\delta N$ formalism.  Here we assume that
the inflaton oscillates under a quadratic potential after inflation and
hence the energy density of the inflaton behaves like a matter during
its oscillation. The homogeneous background equations during reheating
era are given by
\begin{eqnarray}
&&{d\rho_\phi \over dN} + 3\rho_\phi = - {\Gamma \over H}\rho_\phi~, \\ 
&&{d\rho_r \over dN} + 4\rho_r = {\Gamma \over H}\rho_\phi~,\\
&& H^2 = {1 \over 3}\left( \rho_\phi + \rho_r\right)~,
\end{eqnarray}
where $\rho_\phi$ and $\rho_r$ are energy densities of the inflaton
and radiation, respectively. $\Gamma$ is the decay rate of the
inflaton into radiation.  In the modulated reheating scenario, the
decay rate of the inflaton depends on a light scalar field (so-called
modulus) $\sigma$. Thus $\Gamma$ can fluctuate due to fluctuations of
the modulus.  The $e$-folding number $N$ during reheating era can be
given by
\begin{eqnarray} 
N(t_f,t_i) = \ln \left( {a(t_f) \over  a(t_i)}\right)~,
\end{eqnarray}
where $a(t)$ is the scale factor. The initial time $t_i$ and the
final time $t_f$ are respectively taken to be the time at the end of
inflation and some time after the completion of reheating.  Under the
sudden decay approximation, the inflaton decays into the radiation at
$H=\Gamma$, suddenly.  Then, one can rewrite the $e$-folding number as
\begin{eqnarray}
N(t_f,t_i) 
= \ln \left( {a(t_{\rm dec}) \over a(t_i)} \right) + \ln \left( {a(t_f) \over a(t_{\rm dec})}\right)~,
\end{eqnarray}
where $t=t_{\rm dec}$ is the time when $H = \Gamma$.  From $t=t_i$ to
$t=t_{\rm dec}$,  the Universe is dominated by the inflaton behaving
like matter and then becomes dominated by radiation after
$t=t_{\rm dec}$.  Hence, we have
\begin{eqnarray}
N(t_f,t_i) = -{2 \over 3}\ln \left( {\Gamma \over H(t_i)}\right)
-{1 \over 2}\ln \left( {H(t_f) \over \Gamma} \right)~,
\label{efoldreheat}
\end{eqnarray} 
where we have used $H \propto a^{-3/2}$ during matter dominated era and
$H \propto a^{-2}$ during radiation dominated era.  In the  $\delta N$
formalism, the curvature perturbations can be generated from fluctuations
of the decay rate originating from those fluctuations of the modulus $\delta
\sigma_*$,
\begin{eqnarray}
\zeta(t_f) \simeq \delta N 
= N_\sigma \delta \sigma_* + {1 \over 2}N_{\sigma\sigma}\delta \sigma^2_*~,
\label{curvature}
\end{eqnarray}
where
\begin{eqnarray}
&&N_\sigma = {\partial N \over \partial \Gamma} \Gamma'(\sigma)~, \\
&&N_{\sigma\sigma} = {\partial N \over \partial \Gamma} \Gamma''(\sigma)
+{\partial^2 N \over \partial \Gamma^2}\left(\Gamma'(\sigma)\right)^2~.
\end{eqnarray}
From Eq.~(\ref{efoldreheat}), we find
\begin{eqnarray}
&&{\partial N \over \partial \Gamma} = -{1 \over 6}{1 \over \Gamma}~,\label{firstgamma}\\
&&{\partial^2 N \over \partial \Gamma^2} = {1 \over 6}{1 \over \Gamma^2}~.\label{secondgamma}
\end{eqnarray}
Using these expressions, we can evaluate the power spectrum of the
curvature perturbations and also the non-linearity parameter $f_{\rm
  NL}$ in the modulated reheating scenario given by Eq.~(\ref{NLP}).

%%%%%%%%%%
\subsection{Gravitino DM isocurvature fluctuations}\label{sec:GDMmod}
%%%%%%%%%%

Here we discuss isocurvature fluctuations from gravitino DM in the
modulated reheating scenario\footnote{
The same discussion also
applies to axino DM which are produced by thermal scattering
during reheating \cite{Rajagopal:1990yx}.
}. There are two major ways to
produce gravitino DM\footnote{
  Gravitinos can also be produced by the decay of the
  next-to-the-lightest supersymmetric particle (NLSP). However, the
  detailed calculations show that the constraints from BBN are severer
  than those from the overclosure of the universe, irrespective of the kind of
  particle of NLSP as long as it is the MSSM particle
  \cite{Kawasaki:2008qe}. Therefore, a scenario with (most) DM being
  gravitinos can be realized only for the following two cases.
}. One is from the scattering of particles in the thermal plasma and their
relic abundance is evaluated as \cite{Ellis:1984eq},
\begin{equation}
  \frac{n_{3/2}}{s} \simeq 10^{-12}\sum_i g_i^2
              \left(1 + \frac{m_{Gi}^2}{3 m_{3/2}^2}\right)
              \left(\frac{T_R}{10^{10}~{\rm GeV}}\right),
\label{thermalgdm}
\end{equation}
where $T_R$ is the reheating temperature, $m_{3/2}$ is the gravitino
mass, $m_{Gi}$ is the gaugino masses for $i$-th generation, and $g_i$
is the gauge coupling. Notice that the relic abundance is proportional
to the reheating temperature $T_R$. The other way is to produce
gravitinos non-thermally from the decay of some heavy scalar field
such as the inflaton or moduli \cite{Nakamura:2006uc}. In this case,
the yield can be written as
\begin{equation}
\label{eq:Yp_NT}
  \frac{n_{3/2}}{s} = \frac{3}{2} B_{3/2} \frac{T_R}{m_{3/2}},
\end{equation}
where $B_{3/2}$ is the branching ratio of the decay into gravitinos.
When gravitinos are produced from the jets, $B_{3/2}$ should be
understood as those including its multiplicity.  Since the reheating
temperature is related to the decay rate of the inflaton as $\Gamma
\propto T_R^2$, then $B_{3/2} \propto T_R^{-2}$, we have the
$T_R$-dependence of the $n_{3/2} / s$, in the case of non-thermal
production, as
\begin{equation}
%\label{eq:}
 \frac{n_{3/2}}{s} \propto \frac{1}{T_R}.
\label{decaygdm}
\end{equation}
In the modulated reheating scenario, the reheating temperature
fluctuates in space due to the fluctuations of the modulus field
$\sigma$, thus gravitino DM isocurvature fluctuations can be
generated as
\begin{equation}
S_{3/2} = {\delta (n_{3/2} / s) \over n_{3/2} / s} = \pm {\delta T_R \over T_R}~,
\end{equation}
where the positive and negative signs correspond to the cases
with thermal and non-thermal productions, respectively. Hence, we can
easily find that if the gravitinos constitute DM in the Universe,
isocurvature fluctuations are generated in both cases.

From current observations such as CMB, the magnitude of isocurvature
fluctuations is strongly constrained.  Since $T_R$ is proportional to
$\Gamma^{1/2}$, we find $\delta T_R/T_R = \delta\Gamma/(2 \Gamma)$.
Then using the curvature perturbations given by Eq.~(\ref{curvature}),
$S_{3/2}$ can be related to $\zeta$ as
\begin{equation}
  S_{3/2} =  \pm \frac{\delta T_R}{T_R}
               = \pm {\delta \Gamma \over 2\Gamma} \simeq  \mp 3 \zeta.
               \label{eq:iso}
\end{equation}
Thus we have $S_{3/2}/ \zeta \simeq \mp 3$, which is fully correlated to
the curvature perturbations and already contradicts with the current
observations \cite{Takahashi:2009dr}.  In fact, too large isocurvature
fluctuations can also be generated when we consider the
production of gravitino DM in the curvaton mechanism. Since the modulated reheating
and the curvaton scenarios are the major mechanisms of generating large
non-Gaussianity, if the primordial curvature fluctuations are found to be highly
non-Gaussian in the future, gravitino DM scenarios may be
disfavored because of too large isocurvature fluctuations
\cite{Takahashi:2009dr}.  However, when the curvature fluctuations from
the inflaton also contribute to today's density fluctuations, which we
call a mixed scenario, the fraction of isocurvature fluctuations would
be diluted.  Thus in such a case, the DM scenario discussed above
may be liberated. On the other hand, non-Gaussianity, which can be
generated from fluctuations of the modulus or curvaton, would also be
reduced.  Thus it is interesting to investigate how large
non-Gaussianity can be generated without conflicting the constraint on
isocurvature fluctuations in such mixed scenarios of the modulated
reheating and the curvaton.  In the next section, we discuss the mixed
scenario of the modulated reheating.  Then in Section \ref{sec:curv}, we
consider such a scenario in the framework of the curvaton.

%%%%%%%%%%%%%%%%%%%%%%%%%%%%%%%%%%%%%%%%%%
\section{Mixed modulated reheating scenario}\label{sec:mixmod}
%%%%%%%%%%%%%%%%%%%%%%%%%%%%%%%%%%%%%%%%%%

In this section, we discuss the gravitino DM isocurvature fluctuations
and non-Gaussianity in the mixed modulated reheating scenario where
fluctuations of the inflaton also contribute to the curvature
perturbations as well as those from the modulated reheating.

%%%%%%%%%%%%%%
\subsection{Non-Gaussianity}
%%%%%%%%%%%%%%

In the $\delta N$ formalism, the curvature perturbations in the mixed case are,
up to the second order, given by
\begin{eqnarray}
&&\zeta = \zeta_{(\phi)} + \zeta_{(\sigma)}~,
\end{eqnarray}
where $\zeta_{(\phi)}$ and $\zeta_{(\sigma)}$ represent the curvature
perturbations originating from the inflaton and another scalar field
which is assumed to be the modulus here,  respectively. They are written as
\begin{eqnarray}
&&\zeta_{(\phi)} \equiv N_\phi \delta \phi_*, \\
&& \zeta_{(\sigma)} \equiv N_\sigma \delta \sigma_* + {1 \over 2}N_{\sigma\sigma}\delta \sigma_*^2
~.
\end{eqnarray}
Here we have neglected the non-linearity coming from the inflaton
fluctuations since it is of the order of slow-roll parameters and very
small. Notice that $\zeta_{(\phi)}$ or $\zeta_{(\sigma)}$ is
different from $\zeta_{\phi}$ or $\zeta_{\sigma}$, which is often seen
in the literatures.  The former represents the contribution to the
total curvature perturbation  coming from the fluctuations of
$\phi$ or $\sigma$ while the latter is the curvature perturbation on
the slices of uniform density $\rho_{\phi}$ or $\rho_{\sigma}$. Then,
the power spectrum is given by
\begin{eqnarray}
&&P_\zeta (k) = P_\phi (k) + P_\sigma (k)~,
\end{eqnarray}
where $P_\phi$ and $P_\sigma$ are power spectra which are defined by
\begin{eqnarray}
&&\langle \zeta_{(\phi)}(\vec{k}) \zeta_{(\phi)}(\vec{k}')\rangle 
= (2\pi)^3 \delta^{(3)}(\vec{k}+\vec{k}') P_\phi (k)~,\\
&&\langle \zeta_{(\sigma)}(\vec{k}) \zeta_{(\sigma)}(\vec{k}')\rangle 
=(2\pi)^3 \delta^{(3)}(\vec{k}+\vec{k}') P_\sigma (k)~.
\end{eqnarray}
Here we have assumed that $\phi$ and $\sigma$ are uncorrelated.

For the discussion later, we define the ratio of the power spectra
between $P_\sigma$ and $P_\phi$ at some reference scale $k_0$ and
express them as
\begin{eqnarray}
R \equiv \frac{P_\sigma}{P_\phi} 
= \frac{N_\sigma^2}{N_\phi^2} 
\left[ 1 + \frac{N_{\sigma\sigma}^2}{N_\sigma^2} \mathcal{P}_\delta \ln
 (k L) \right]~,
 \label{ratio}
\end{eqnarray}
with 
\begin{eqnarray}
N_\phi^2 &=& \left({V \over V'}\right)^2 = {1 \over 2\epsilon}~,\label{inflaton} \\
{N_\sigma^2 } &\! = \!& { 1 \over 36}\left( {\Gamma' \over \Gamma} \right)^2~,
\label{first}\\
{N_{\sigma\sigma} \over N_\sigma^2} &\! = \!&
 6 \left[ 1 - {{\Gamma'' / \Gamma} \over \left( \Gamma' / \Gamma \right)^2} \right]~.\label{second}
\end{eqnarray}
Here for simplicity we have considered the standard slow-roll
inflation model and $\epsilon$ is so-called a slow-roll parameter.
With this definition, the limits of $R\rightarrow 0$ and $\infty$
correspond to the pure inflaton and pure modulus cases, respectively.  By
using $R$, the non-linearity parameter $f_{\rm NL}$ given by
Eq.~(\ref{NLP}) can be rewritten as
\begin{eqnarray}
{6 \over 5}f_{\rm NL} = {R \over \left( 1 + R \right)^2}
{N_{\sigma\sigma} \over N_\phi^2}~.\label{fnl}
\end{eqnarray}

In principle, the term with $N_{\sigma\sigma\sigma}$ can also arise in
$R$\footnote{
  In fact, the term with $\Gamma^{\prime\prime\prime} = d^3
  \Gamma(\sigma) / d\sigma^3$ can also appear. However, in an explicit
  model we consider in the following, such higher order derivatives
  vanish. Thus we neglect such a term.
}. However, by a simple inspection of such a term, we can see that
such higher order derivative terms can be neglected. If we take into
account the contribution from the third order term in $\delta \sigma$,
$R$ can be expressed as
\begin{eqnarray}
R = {N_\sigma^2 \over N_\phi^2}
\left[ 1 + \left({N_{\sigma\sigma}^2 \over N_\sigma^2} + {N_{\sigma\sigma\sigma} \over N_\sigma}\right)
{\cal P}_\delta \ln (k_m L ) \right]~.
\end{eqnarray}
Now we compare $N_{\sigma\sigma}^2/N_\sigma^2$ with
$N_{\sigma\sigma\sigma}/N_\sigma$ and show that the latter is much
smaller than the former, in particular,  when $f_{\rm NL}$ is large.  From
Eqs.~(\ref{first}) and (\ref{second}), we obtain
\begin{eqnarray}
N_{\sigma\sigma\sigma} \simeq -{1 \over 6}
\left[
2\left( {\Gamma' \over \Gamma} \right)^3 
- 3 \left( {\Gamma' \over \Gamma}\right)\left( {\Gamma'' \over \Gamma} \right)
\right]~.
\label{eq:third}
\end{eqnarray}
Thus the ratio of these combinations is given by
\begin{eqnarray}
{N_{\sigma\sigma\sigma}/ N_{\sigma} \over N_{\sigma\sigma}^2 / N_\sigma^2}
= \frac{2 - 3 \displaystyle{\Gamma'' / \Gamma \over \left( {\Gamma' / \Gamma} \right)^2}}
{ \left[ 1 - \displaystyle{\Gamma'' / \Gamma \over \left( {\Gamma' / \Gamma} \right)^2} \right]^2}~.
\label{eq:thirdsecond}
\end{eqnarray}
From Eq.~(\ref{fnl}) we can easily find that at least we need
$N_{\sigma\sigma} / N_\sigma^2 \gg 1$ in order to realize large
$f_{\rm NL}$ and then this condition corresponds to
\begin{eqnarray}
\left|
{\Gamma'' / \Gamma \over \left( {\Gamma' / \Gamma} \right)^2}
\right| \gg 1~.
\label{eq:condition}
\end{eqnarray}
Under this condition Eq.~(\ref{eq:thirdsecond}) is approximately given
by
\begin{eqnarray}
\left|
{N_{\sigma\sigma\sigma}/ N_{\sigma} \over N_{\sigma\sigma}^2 / N_\sigma^2}
\right|
 \sim \left[ \frac{\Gamma'' / \Gamma}{\left( {\Gamma' / \Gamma} \right)^2} \right]^{-1}
\ll 1 ~.
\end{eqnarray}
Hence as far as we consider the case with $f_{\rm NL} \gg \mathcal{O}(1)$, the term with
$N_{\sigma\sigma\sigma}$ in the one-loop correction is negligible.

Similarly, we can easily confirm that higher order terms like
$N_{\sigma\sigma\sigma\sigma}$ and $N_{\sigma\sigma\sigma\sigma\sigma}$,
which can arise in the one-loop correction term of bispectrum and
trispectrum, are also negligible.

%%%%%%%%%%%%%
\subsection{Gravitino DM isocurvature fluctuations}
%%%%%%%%%%%%%

Here, we discuss the gravitino DM isocurvature fluctuations
in the mixed modulated reheating  scenario.  Gravitino DM isocurvature fluctuations are given by
\begin{eqnarray}
S_{3/2} &\! = \!& {1 \over 2}{\Gamma' \over \Gamma}\delta \sigma_*
+ {1 \over 2}\left[{1 \over 2}{\Gamma'' \over \Gamma}
 - {1 \over 4}\left( {\Gamma' \over \Gamma}\right)^2 \right]\delta \sigma_*^2 \nonumber\\
&\! = \!& -3 \left[ N_\sigma \delta \sigma_* 
+ {1 \over 2}N_{\sigma\sigma}\left( 1 - 3 {N_\sigma^2 \over N_{\sigma\sigma}} \right) \delta \sigma_*^2 \right] 
 ~,\label{eq:gdmiso}
\end{eqnarray}
which are correlated with the adiabatic fluctuations as $S_{3/2} \simeq - 3
\zeta_\sigma$ as discussed in the previous section\footnote{
  In fact, the relation between the isocurvature and 
  adiabatic fluctuations slightly deviates from $S_{3/2} = - 3
  \zeta_\sigma$ due to the non-linear terms. However, in order to realize
  large non-Gaussianity, the condition given by Eq.~(\ref{eq:condition})
  must be satisfied. Hence, by neglecting $N_\sigma^2 /
  N_{\sigma\sigma}$ in the non-linear term of $S_{3/2}$ we approximately
  obtain $S_{3/2} \simeq - 3\zeta_\sigma$.
}. Since here we consider a mixed scenario where the curvature
fluctuations can also be generated from the inflaton, only some
fraction of the isocurvature perturbations is correlated with the
(total) curvature perturbations.
Then, $S_{3/2}$ can be divided into the following two parts,
\begin{eqnarray}
 S_{3/2} &=& S_{\rm corr} + S_{\rm uncorr}, \nonumber \\
 S_{\rm corr} &=& -3 
  \frac{P_{\sigma}}{P_{\zeta}}
  \left( \zeta_{(\phi)} + \zeta_{(\sigma)} \right)
  = - \frac{3R}{1+R} \zeta, \nonumber \\
 S_{\rm uncorr} &=& 
  \frac{3R}{1+R}
  \left( \zeta_{(\phi)} - \frac{P_{\phi}}{P_{\sigma}} \zeta_{(\sigma)}
  \right)
  = \frac{3}{1+R}
  \left( R \zeta_{(\phi)} - \zeta_{(\sigma)} \right), 
   \label{eq:divided}
\end{eqnarray}
where the first part is denoted as $S_{\rm corr}$ and the second as
$S_{\rm uncorr}$. Using the expression~(\ref{ratio}), we have the
following relations,
\begin{eqnarray}
P_{S_{\rm corr}} = {R \over 1+R}P_{S_{3/2}}~,~
P_{S_{\rm uncorr}} = {1 \over 1+R}P_{S_{3/2}}~.
\label{eq:coruncor}
\end{eqnarray}
To express the size of the contribution from isocurvature
fluctuations, we define the ratio of the power spectrum relative to
the total one as
\begin{eqnarray}
{\alpha_{\rm corr}} \equiv {P_{S_{\rm corr}} \over P_{\zeta}+P_{S_{\rm corr}}}~, \\
{\alpha_{\rm uncorr}} \equiv {P_{S_{\rm uncorr}} \over P_{\zeta}+P_{S_{\rm uncorr}}}~,
\end{eqnarray}
for the correlated and uncorrelated parts, respectively. Following the
notation of Ref.~\cite{Komatsu:2008hk}, we define the cross-correlation
coefficient $\beta$ as
\begin{eqnarray}
\beta \equiv - \frac{P_{S_{\rm corr}\zeta}}{\sqrt{P_\zeta P_{S_{\rm corr}}}}~,
\label{eq:crosscorr}
\end{eqnarray}
where $P_{S_{\rm corr}\zeta}$ denotes the cross-correlation power spectrum defined by
\begin{eqnarray}
\langle S_{{\rm corr} \vec{k}_1} \zeta_{\vec{k}_2}\rangle
\equiv  (2\pi)^3 \delta^{(3)}\left( \vec{k}_1 + \vec{k}_2\right)P_{S_{\rm corr}\zeta}(k_1)~.
\end{eqnarray}

Hence the correlation coefficient is $\beta = +1 (-1)$ for the
thermally (non-thermally) produced gravitino DM in the modulated
reheating scenario.

As already mentioned, the size of isocurvature fluctuations is now
severely constrained by observations of CMB and so
on. In fact, in the present model, correlated and uncorrelated
isocurvature perturbations coexist, thus we need to take into account
both contributions simultaneously to obtain observational constraints
on $\alpha_{\rm corr}$ and $\alpha_{\rm uncorr}$.  However, such
analysis is not available in the literatures. Hence as reference
values, we adopt the constraints on $\alpha_{\rm corr}$ and
$\alpha_{\rm uncorr}$ obtained separately from recent WMAP5 results:
$\alpha_{\rm corr} < 0.011$ and $\alpha_{\rm uncorr} < 0.16$ at 95 \%
C.L. from the WMAP-only analysis \cite{Komatsu:2008hk}\footnote{
  In fact, the constraint for correlated isocurvature fluctuations
  here is obtained for $\beta = -1$. However, it is expected that the
  sign of the correlation does not affect the constraint on the size
  much. Thus we refer this value regardless of the sign of $\beta$ as
  a reference value.
}.

%%%%%%%%%%%%%%%%%
\subsection{A Simple  Model} \label{subsec:toymodel}
%%%%%%%%%%%%%%%%%

Now let us work on some explicit model of the modulated reheating
scenario.  We consider the following interaction between an inflaton
$\phi$ and a fermion $\psi$:
\begin{eqnarray}
{\cal L}_{\rm int} \sim -g({\sigma}) \phi \overline{\psi} \psi~,
\end{eqnarray}
where the coupling constant $g$ depends on a modulus field $\sigma$.
We assume that $g(\sigma)$ can be written as
\begin{eqnarray}
g(\sigma) \simeq g_0 \left[ 1 + g_1 \left( {\sigma \over M}\right) + g_2 \left( {\sigma \over M} \right)^2 \right]~,
\end{eqnarray}
where $M$ is some energy scale, $g_1$ and $g_2$ are some coefficients
and $|\sigma| / M < 1$ is assumed.  The decay rate of the inflaton
through this interaction is given by
\begin{eqnarray}
\Gamma \sim {g^2 \over 8\pi}m_\phi~,
\end{eqnarray}
which implies that the form of the decay rate of inflaton is 
\begin{eqnarray}
\Gamma = \Gamma_0 \left[ 1 + A \left( {\sigma \over M} \right) 
+ B \left( {\sigma \over M}\right)^2\right]~,
\label{toymodel}
\end{eqnarray}
where $A$ and $B$ are some coefficients.  Substituting
Eq.~(\ref{toymodel}) to Eqs.~(\ref{first}) and (\ref{second}), we
obtain
\begin{eqnarray}
{N_\sigma^2 } = 
{1 \over 36 M^2}
\left[ {A + 2B \left( \sigma / M \right) \over 1 
+ A \left( \sigma / M \right) + B \left( \sigma / M\right)^2 } \right]^2~,
\label{firsttoy}
\end{eqnarray}
and
\begin{eqnarray}
{N_{\sigma\sigma} } =
{1\over 6M^2} \left[  \left({A + 2B \left( \sigma / M \right) \over 1 
+ A \left( \sigma / M \right) + B \left( \sigma / M\right)^2 } \right)^2 
- 
{2B 
\over
1 
+ A \left( \sigma / M \right) + B \left( \sigma / M\right)^2} \right]~.
\label{secondtoy}
\end{eqnarray}

Now we discuss the non-linearity parameter $f_{\rm NL}$ and the size of
isocurvature fluctuations in this model.  There are four parameters in
the model; $M$, $\sigma/M$, $A$ and $B$.  Here, we will show the results
for $f_{\rm NL}$  and the isocurvature fraction by considering the limit of 
$|\sigma |/M  \ll 1$ or $A \rightarrow 0$.

\begin{figure}[htbp]
  \begin{center}
    \includegraphics[keepaspectratio=true,height=50mm]{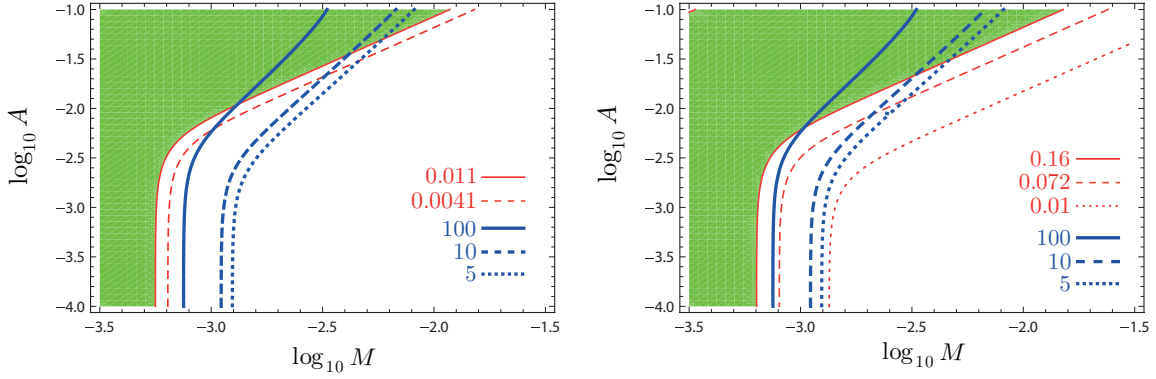}
  \end{center}
  \caption{(color online) In the left (right) panel, contours of $f_{\rm NL}$ and ${\alpha}_{\rm corr}$ 
  ($\alpha_{\rm uncorr}$) in the
    $M$--$A$ plane are shown.  We have fixed as $B=-1.0$
    and assumed $\left| \sigma \right| / M \ll 1$.  Red thin lines in the left panel show
    contours of ${\alpha}_{{\rm corr}}=0.011$ (solid line) and
    $0.0041$ (dashed line). In the right panel, the lines are shown for
    ${\alpha}_{{\rm uncorr}}=0.16$ (solid line), $0.072$
    (dashed line) and $0.01$(dotted line).  Green shaded regions are
    constrained by the current observational limit for the
    isocurvature fractions from WMAP5-only analysis.
    Blue thick lines show contours of $f_{\rm NL} = 5$ (dotted line),
    $10$ (dashed line) and $100$ (solid line).  }
  \label{fig:alphaM_cor_un.eps}
\end{figure}

\begin{figure}[htbp]
  \begin{center}
    \includegraphics[keepaspectratio=true,height=50mm]{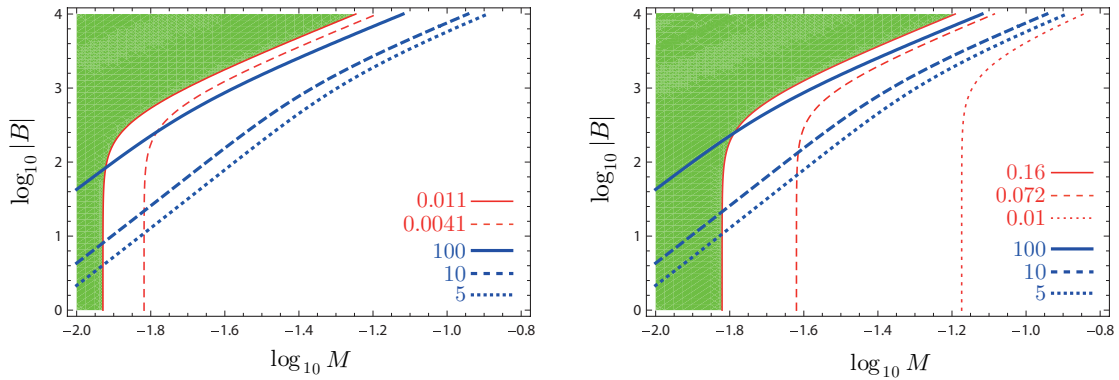}
  \end{center}
  \caption{(color online) In the left (right) panel, contours of $f_{\rm NL}$ and ${\alpha}_{\rm corr}$ 
  ($\alpha_{\rm uncorr}$) in the
    $M$--$|B|$ plane are shown.  We have fixed as $A=1.0$
    and assumed $\left| \sigma \right| / M \ll 1$. }
  \label{fig:betaM_cor_un.eps}
\end{figure}

\begin{figure}[htbp]
  \begin{center}
    \includegraphics[keepaspectratio=true,height=50mm]{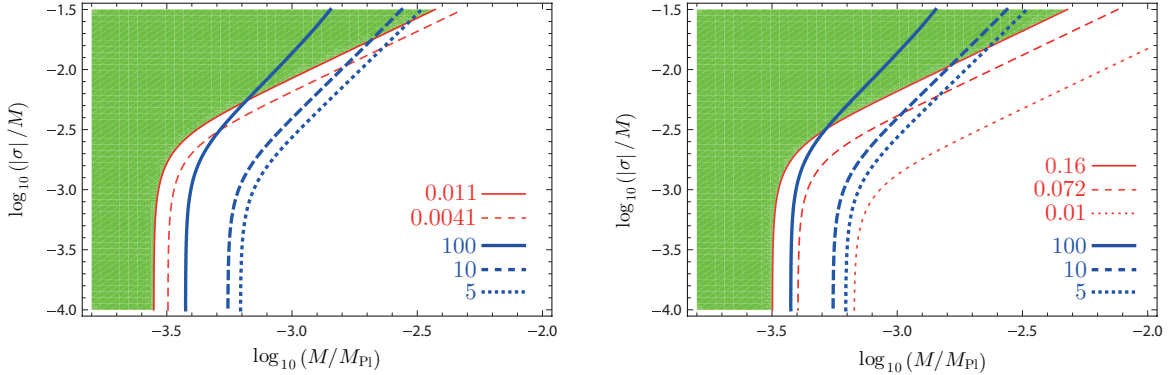}
  \end{center}
  \caption{(color online) In the left (right) panel, contours of $f_{\rm NL}$ and ${\alpha}_{\rm corr}$ 
  ($\alpha_{\rm uncorr}$) in the
    $M$--$\left| \sigma \right| / M$ plane are shown.  We have fixed as
    $B=-1.0$ and $A=0$.  }
  \label{fig:alphazero_cor_un.eps}
\end{figure}

In the case of $\left| \sigma \right| / M \ll 1$, $N_\sigma^2$ in
Eq.~(\ref{firsttoy}) and $N_{\sigma\sigma}$ in Eq.~(\ref{secondtoy})
are respectively reduced as
\begin{equation}
{N_\sigma^2} \simeq
\left({A \over 6 M}\right)^2
~,
~~~~~~
{N_{\sigma\sigma}}
\simeq
{1 \over 6M^2} \left( A^2 - 2B \right)
~.
\end{equation}
On the other hand, for the limit of $A \to 0$, these are written as
\begin{equation}
{N_\sigma^2 } \simeq
{1 \over 36M^2}
\left[
{2B \left( \sigma / M \right) \over 1+B \left( \sigma / M \right)^2}
\right]^2~,
~~~
{N_{\sigma\sigma} }
\simeq
{1 \over 6M^2}\left[
\left(
{2B \left( \sigma / M \right) \over 1+B \left( \sigma / M \right)^2}
\right)^2 
- { 2B \over 1+ B \left( \sigma / M \right)^2}
\right]~.
\end{equation}

In Fig.~\ref{fig:alphaM_cor_un.eps}, we show contours of $f_{\rm NL}$
along with ${\alpha}_{\rm corr}$ (left panel) and $\alpha_{\rm uncorr}$
(right panel) in the $M$--$A$ plane.  We have fixed the value of $B$ as
$B=-1.0$ and assumed $\left| \sigma \right| / M \ll 1$.  In
Fig.~\ref{fig:betaM_cor_un.eps}, we show the same but in the $M$--$|B|$
plane.  In the figure, we have fixed as $A=1.0$ and assumed $\left| \sigma \right| 
/ M \ll 1$.  In Fig.~\ref{fig:alphazero_cor_un.eps}, we have fixed the value
of model parameters as $B=-1.0$ and $A=0$, then show contours of $f_{\rm
NL}$ as well as ${\alpha}_{\rm corr}$ and $\alpha_{\rm uncorr}$ in the
$M$--$ \left| \sigma \right| / M$ plane.  Regarding the slow-roll parameter $\epsilon$
for the inflaton, we take $\epsilon = 0.01$ in the following analyses.  In all
figures, red thin lines in the left panel correspond to contours of
${\alpha}_{{\rm corr}}=0.011$ (solid line) and $0.0041$ (dashed line).
These numbers correspond to the 95 \% C.L. limit from WMAP5 only and
WMAP5+BAO+SN analyses, respectively.  In the right panel, they
correspond to contours of ${\alpha}_{{\rm uncorr}}=0.16$ (solid line)
and $0.072$ (dashed line). These numbers again correspond to the 95 \%
C.L. limit from WMAP5 only and WMAP5+BAO+SN analyses, respectively. In
the near future, we will have a more stringent limit from PLANCK
\cite{Enqvist:1999vt,planck:2006uk,Bucher:2000kb} and its projected
limit on uncorrelated isocurvature fluctuations will be $\alpha_{\rm
uncorr} < 0.01$, thus we also show the contour of ${\alpha}_{{\rm
uncorr}}=0.01$ with dotted line.  Green shaded regions are constrained
by the current observational limit for the isocurvature fractions from
WMAP-only analysis.  Blue thick lines show contours of $f_{\rm NL} = 5$
(dotted line), $10$ (dashed line) and $100$ (solid line).

From these figures, we can see that large values of $f_{\rm NL} \sim
(10-100)$ can still be realized without conflicting the current constraint
on isocurvature fluctuations.  In particular, 
when the one-loop correction term dominates, the following simple 
relation holds between $f_{\rm NL}$ and
$\alpha_{\rm uncorr}$:
\begin{eqnarray}
f_{\rm NL} \simeq 3 \times \left( {\alpha_{\rm uncorr} \over 0.01} \right)^{3/2}~.
\label{eq:uncorfnl}
\end{eqnarray}
In Figs.~\ref{fig:alphaM_cor_un.eps} and~\ref{fig:alphazero_cor_un.eps},
the regions where the one-loop term dominates 
correspond to downside of the figures and in
Fig.~\ref{fig:betaM_cor_un.eps},  it corresponds to upside of the
figure.
In these parameter regions, the present limit on the
isocurvature fluctuations still allows relatively large non-Gaussianity
in the scenario.

Here it should be mentioned that the isocurvature constraint is severer
for uncorrelated one.  The reason is as follows.  Here we consider the
case where the adiabatic curvature perturbations are mainly generated
from fluctuations of the inflaton field, which corresponds to the
case with $R \ll 1$.  On the other hand, the gravitino DM isocurvature
fluctuations are produced from fluctuations of the modulus field as
given in Eq.~(\ref{eq:gdmiso}).  Hence, in the small $R$ region, the
gravitino DM isocurvature fluctuations have mostly become uncorrelated
type. This fact can be also understood by noting Eq.~(\ref{eq:coruncor}).  In
the near future, the data from PLANCK will be available and the
constraint on isocurvature fluctuations would be much severer than that
of the current one.  The projected limit on uncorrelated isocurvature
fluctuations would be $\alpha_{\rm uncorr} < 0.01$ at 95 \%
C.L. \cite{Enqvist:1999vt}.  Thus, for reference, we also plot the
corresponding contour in the figures, from which we can find that, if we
obtain such a stringent constraint, large non-Gaussianity cannot be
generated even in the mixed scenario.  Thus if in the future, large
local-type non-Gaussianity is confirmed, but the isocurvature constraint becomes
as severe as that mentioned above, gravitino DM scenario would be
ruled out even if we consider the mixed scenario.

\subsection{Brief comments on the trispectrum}

Before closing this section, we would like to make brief comments on
the trispectrum in the mixed modulated reheating scenario.  The local
type trispectrum can be parameterized with two non-linearity
parameters $g_{\rm NL}$ and $\tau_{\rm NL}$ as \cite{Byrnes:2006vq}
\begin{eqnarray}
\langle \zeta_{\vec{k}_1}\zeta_{\vec{k}_2}\zeta_{\vec{k}_3}\zeta_{\vec{k}_4}\rangle &\! \equiv \!&
\left( 2\pi \right)^3
\biggl[ \tau_{NL}\left( P_\zeta(k_{13})P_\zeta(k_3)P_\zeta(k_4) + 11~{\rm perms.}\right) \nonumber\\
&& \quad\quad\quad
+ {54 \over 25}g_{NL}\left( P_\zeta(k_{2})P_\zeta(k_3)P_\zeta(k_4) + 3~{\rm perms.}\right)
\biggr] \delta^{(3)}
\left( \vec{k}_1 + \vec{k}_2 + \vec{k}_3 + \vec{k}_4\right)~,\nonumber\\
\end{eqnarray}
where $k_{ij} = \left| \vec{k}_i + \vec{k}_j\right|$.  By adopting the
$\delta N$ formalism, in the mixed modulated reheating scenario we
have
\begin{eqnarray}
\tau_{NL} &\! = \!& { R \over \left(1+R\right)^3 }
\left({N_{\sigma\sigma} \over N_\phi^2}\right)^2~, \label{eq:tauNL}
\end{eqnarray}
including the one-loop correction.
From Eqs.~(\ref{fnl}) and (\ref{eq:tauNL}),
\begin{eqnarray}
\label{eq:tauNL_fNL}
\tau_{\rm NL} \simeq \left( {1 + R \over R} \right) f_{\rm NL}^2~.
\end{eqnarray}
Hence, for small $R$ there is a possibility of generating large
$\tau_{\rm NL}$ which may be detected in the future experiments,
without contradicting the current observational constraint on the
isocurvature fluctuations. As discussed above, once we obtain a severe
constraint on isocurvature fluctuations, the value of $f_{\rm NL}$
would be small in the mixed modulated reheating scenario.  However,
it is possible that a non-Gaussian
signature comes from the trispectrum but not from the bispectrum.

%%%%%%%%%%%%%%%%%%%%%%%%%%%%%%%%%%%%
\section{Mixed curvaton scenario}\label{sec:curv}
%%%%%%%%%%%%%%%%%%%%%%%%%%%%%%%%%%%%

Now in this section, we consider CDM isocurvature fluctuations in the
curvaton scenario.  Isocurvature 
fluctuations in the curvaton have been investigated in the literatures
\cite{Moroi:2002rd,Lyth:2002my,Lyth:2003ip,Beltran:2008ei,Moroi:2008nn}.
Here we investigate this issue in the framework of 
a mixed scenario and focusing on how large non-Gaussianity 
can be produced without conflicting with the constraints on 
isocurvature fluctuations.
Notice that the discussions given in this section apply not only to
gravitino (axino) DM but also a generic CDM,
although we focused on such DM candidates in the previous section.  
As pointed out in
\cite{Lyth:2003ip,Takahashi:2009dr}, when the number density of CDM
freezes before the curvaton decay, too large isocurvature fluctuations
are generated and excluded by cosmological observations. However, this
conclusion is valid for the original curvaton scenario in which
fluctuations from the curvaton are only responsible for density
fluctuations today.  But, in general, fluctuations of the inflaton
also contribute in addition to those from the
curvaton. Such a mixed scenario has been extensively studied for the
adiabatic fluctuations in \cite{Langlois:2004nn,Moroi:2005kz,Moroi:2005np,Ichikawa:2008iq,Langlois:2008vk} and for
baryon isocurvature fluctuations \cite{Moroi:2008nn}.  Here we consider
CDM isocurvature fluctuations in the mixed curvaton scenario for the
cases with CDM being produced from the decay of the inflaton and/or the
curvaton, paying particular attention to how large non-Gaussianity can be
without conflicting with the isocurvature constraint.

In the same way as in the mixed modulated scenario, under the sudden
decay approximation, the adiabatic curvature perturbations on the
uniform (total) energy density hypersurface are analytically given
by~\cite{Langlois:2008vk}
\begin{eqnarray}
&&\zeta = \zeta_{(\phi)} + \zeta_{(\sigma)}~\nonumber\\
&&\zeta_{(\phi)} = N_\phi \delta \phi_*~,\nonumber\\
&& \zeta_{(\sigma)} = N_\sigma \delta \sigma_* + N_{\sigma\sigma}\delta \sigma_*^2
= 
{2 \over 3} f_{\rm dec} \frac{\delta \sigma_*}{\sigma_*}
+ \frac13 f_{\rm dec}
\left( 1 - \frac{4}{3}f_{\rm dec} - \frac{2}{3}f_{\rm dec}^2\right)
\left( \frac{\delta \sigma_*}{\sigma_*} \right)^2~,
\label{eq:adiabatic}
\end{eqnarray}
where $\phi$ and $\sigma$ denote the inflaton and the curvaton,
respectively, and $\sigma$ is taken to be positive without loss of generality\footnote{
The case where the second term dominates the
linear term in $\zeta_{(\sigma)}$ corresponds to the ``ungaussiton"
scenario \cite{Linde:1996gt,Boubekeur:2005fj,Suyama:2008nt}. 
Our analysis includes such a scenario automatically.
}.
We have neglected the non-linear
part of fluctuations of the inflaton since they are very small.
Here $f_{\rm dec}$ is defined by
\begin{eqnarray}
f_{\rm dec} = \frac{3\rho_\sigma}{4\rho_r + 3 \rho_\sigma}\biggr|_{t=t_{\rm dec}}~,
\end{eqnarray}
where $\rho_\sigma$ and $\rho_r$ are respectively energy densities
of the curvaton field and radiation. $t=t_{\rm dec}$ is the
time at the curvaton decay.  For large non-Gaussianity, at least we
need
\begin{eqnarray}
\frac{N_{\sigma\sigma}}{N_\sigma^2}
= 
\frac{3}{4}\frac{1}{f_{\rm dec}}
\left( 1 - \frac{4}{3}f_{\rm dec} - \frac{2}{3}f_{\rm dec}^2\right) \gg 1~,
\end{eqnarray}
which leads to $f_{\rm dec} \ll 1$. Hereinafter, we consider the case
where this condition is satisfied.

After the curvaton decay, the CDM isocurvature fluctuations are expressed as
\begin{eqnarray}
S_{\rm CDM} = 3\left( \zeta_{\rm CDM} - \zeta \right)~,
\end{eqnarray}
where $\zeta$ is given by Eq.~(\ref{eq:adiabatic}) and $\zeta_{\rm CDM}$
is the curvature perturbation on the uniform CDM energy
density hypersurface.

Regarding the production of CDM, one can consider two cases: dominant
residual CDM is generated from the decay of the inflaton or the
curvaton. In the following, we consider each case separately.

%%%%%%%%%%%%%%
\subsection{Case with CDM from the inflaton decay}
%%%%%%%%%%%%%%

First, let us consider the case where dominant residual CDM is
generated from the decay of inflaton, which implies that CDM has the
same fluctuations of the inflaton, that is, $\zeta_{\rm CDM} =
\zeta_\phi$ which denotes the curvature perturbation on uniform inflaton
energy density hypersurface.
In the case where $f_{\rm dec} \ll 1$, 
we can consider
$\zeta_\phi \simeq \zeta_{(\phi)}$ and hence
$\zeta_{\rm CDM} \simeq \zeta_{(\phi)}$.
Then, we obtain~\cite{Langlois:2008vk}
\begin{equation}
\label{eq:curvaton_iso_inf}
S_{\rm CDM} = -3 \left( N_\sigma \delta \sigma_* + {1 \over 2}N_{\sigma\sigma} \delta \sigma_*^2 \right)
= - 3 \zeta_{(\sigma)}~.
\end{equation}
Since we are considering a mixed scenario, the isocurvature
fluctuations here should have uncorrelated and correlated parts with
adiabatic fluctuations which originate from both the inflaton and the
curvaton. With the definition of the correlation coefficient of
Eq.~\eqref{eq:crosscorr}, $\beta = 1$ for this case.  Then we separate
isocurvature fluctuations into two parts as done in the previous
section.

\begin{figure}[htbp]
  \begin{center}
    \includegraphics[keepaspectratio=true,height=50mm]{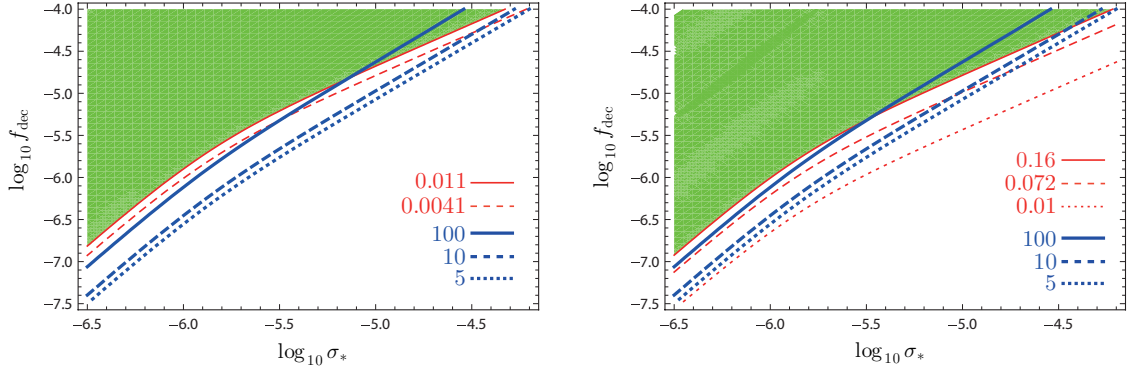}
  \end{center}
  \caption{(color online) Contours of $f_{\rm NL}$ along with ${\alpha}_{\rm corr}$
    (left panel) and $\alpha_{\rm uncorr}$ (right panel) in the
    $\sigma_*$--$f_{\rm dec}$ plane.  Red thin lines in the left panel show
    contours of ${\alpha}_{{\rm corr}}=0.011$ (solid line) and
    $0.0041$ (dashed line).  In the right panel, we show contours
    of ${\alpha}_{{\rm uncorr}}=0.16$ (solid line), $0.072$ (dashed
    line) and $0.01$ (dotted line).  Green shaded regions are
    constrained by the current observational limit on
    isocurvature fractions from WMAP5 only.  Blue thick lines are contours
    of $f_{\rm NL} = 5$ (dotted line), $10$ (dashed line) and $100$
    (solid line).}
  \label{fig:curvinf_cor_un.eps}
\end{figure}

In Fig.~\ref{fig:curvinf_cor_un.eps}, we plot contours of $f_{\rm NL}$
as well as $\alpha_{\rm corr}$ (left panel) and $\alpha_{\rm uncorr}$
(right panel) in the $\sigma_\ast$--$f_{\rm dec}$ plane.  The results
are quite similar to those for the mixed modulated reheating scenario
discussed in the previous section. At the current level of the
constraint on isocurvature fluctuations, relatively large
non-Gaussianity can be generated without conflicting the constraint.
However, once the limit becomes severe as $\alpha_{\rm uncorr} < 0.01$
which is expected in PLANCK, $f_{\rm NL}$ should not be large in this
case.

\subsection{Case with CDM from the curvaton decay}

Next, let us consider the case where dominant CDM component is
generated from the decay of the curvaton, which implies that $\zeta_{\rm
CDM}=\zeta_{\sigma} (\ne \zeta_{(\sigma)} = f_{\rm dec} \zeta_\sigma)$ because $f_{\rm dec} \ll 1$,
where $\zeta_\sigma$ is the curvature perturbation on the uniform curvaton energy density hypersurface. 
Then, we can obtain \cite{Langlois:2008vk}
\begin{equation}
\label{eq:curvaton_iso_curvaton}
S_{\rm CDM} = (1-f_{\rm dec})
  \left[2 \frac{\delta \sigma_*}{\sigma_*}
        - \left(1 + 2f_{\rm dec} + \frac23 f_{\rm dec}^2 \right)
          \left( \frac{\delta \sigma_*}{\sigma_*} \right)^2 
  \right].
\end{equation}
In the case where $f_{\rm dec} \ll 1$, $S_{\rm CDM}$  is approximately given by
\begin{eqnarray}
S_{\rm CDM} \simeq {3 \over f_{\rm dec}}\left[N_\sigma \delta \sigma_*
-{1 \over 2}N_{\sigma\sigma} \delta \sigma_*^2\right]
~,\label{eq:isocurvcurv}
\end{eqnarray}
with
\begin{eqnarray}
N_\sigma = {2 \over 3}{f_{\rm dec} \over \sigma_*}~,~~
N_{\sigma\sigma} =  {2 \over 3}{f_{\rm dec} \over \sigma_*^2}~,
\label{eq:expressionN}
\end{eqnarray}
from Eq.~(\ref{eq:adiabatic}).  Then, the power spectrum of the
isocurvature fluctuations is given by
\begin{eqnarray}
P_{S_{\rm CDM}}(k) &\! \simeq \!& 9 {1 \over f_{\rm dec}^2}
\left[ N_\sigma^2 + N_{\sigma\sigma}^2 {\cal P}_\delta \ln(k L) \right] P_\delta(k) \nonumber\\ 
&\! = \!& 9{1 \over f_{\rm dec}^2}P_{\sigma}(k)~.
\label{eq:poweriso}
\end{eqnarray}
In fact, in this case, 
we can easily find that
CDM isocurvature fluctuations become large even for $f_{\rm NL} = O(1)$. 
From the above equation, the ratio of the power spectrum of the CDM isocurvature
fluctuations to that of the adiabatic (curvature) perturbations is given by 
\begin{eqnarray}
\frac{P_{S_{\rm CDM}}}{P_\zeta} = \frac{9}{f_{\rm dec}^2}\frac{P_{\sigma}}{P_\zeta} = 
\frac{9}{f_{\rm dec}^2} R ~.
\end{eqnarray}
From Eqs.~(\ref{ratio}), we obtain the following inequality:
\begin{eqnarray}
R \ge  \frac{N_\sigma^2}{N_\phi^2} = {8 \over 9}\epsilon {f_{\rm dec}^2 \over \sigma_*^2}~,
\label{eq:ampiso}
\end{eqnarray}
where we have used $N_\phi^2 = \left( 2\epsilon \right)^{-1}$ and Eq.~(\ref{eq:expressionN}).
Hence, we have the inequality given by
\begin{eqnarray}
\frac{P_{S_{\rm CDM}}}{P_\zeta}
\ge  8 {\epsilon \over \sigma_*^2}~.
\label{eq:powerratio}
\end{eqnarray}
On the other hand, from Eqs.~(\ref{fnl}), assuming that $R \ll 1$,
the non-linearity parameter $f_{\rm NL}$ can be written as, 
\begin{eqnarray}
f_{\rm NL} \simeq  R \frac{N_{\sigma\sigma}}{N_\phi^2}
= \frac{4}{3}f_{\rm dec} R \frac{\epsilon}{\sigma_*^2}~.
\label{eq:ampfnl}  
\end{eqnarray}
Combining Eq.~(\ref{eq:ampfnl}) with Eq.~(\ref{eq:powerratio}),
we obtain 
\begin{eqnarray}
\frac{P_{S_{\rm CDM}}}{P_\zeta}
\ge R^{-1} f_{\rm dec}^{-1} f_{\rm NL}.
\end{eqnarray}
When $R \ll 1$ and $f_{\rm dec} < 1$, 
the fraction of the power spectrum of the 
CDM isocurvature fluctuations to that of the adiabatic (curvature) perturbations becomes
much larger than the size of the non-Gaussianity, $P_{S_{\rm CDM}} / P_\zeta \gg f_{\rm NL}$.
Thus, in the case where dominant CDM component is generated from the decay of curvaton,
large non-Gaussianity from the (adiabatic) curvature fluctuations cannot be 
generated without conflicting observational constraint on CDM isocurvature fluctuations
even at the current level.

%%%%%%%%%%%%%%%%%%%%%%%%%%%%%%%%%%%%%
\section{Non-Gaussianity from DM isocurvature fluctuations}\label{sec:iso}
%%%%%%%%%%%%%%%%%%%%%%%%%%%%%%%%%%%%%

So far we have discussed non-Gaussianity of adiabatic fluctuations.
However, isocurvature fluctuations can also produce large
non-Gaussianity, which has been discussed recently in
\cite{Kawasaki:2008sn,Kawasaki:2008pa,Langlois:2008vk,Hikage:2008sk}. Thus
in this section, we investigate non-Gaussianity from isocurvature
fluctuations in our scenario.

First, we consider non-Gaussianity from isocurvature fluctuations in the
modulated reheating scenario. As discussed in Sec.~\ref{sec:mixmod},
the isocurvature fluctuations can be written, up to the second order, as
\begin{eqnarray}
\label{eq:S32}
S_{3/2} = {1 \over 2}{\Gamma' \over \Gamma}\delta \sigma_*
+ {1 \over 2}\left[{1 \over 2}{\Gamma'' \over \Gamma}
 - {1 \over 4}\left( {\Gamma' \over \Gamma}\right)^2 \right]\delta \sigma_*^2~.
\end{eqnarray}
Depending on the explicit form of $\Gamma$ and the value of $\delta
\sigma_\ast$, the discussion on non-Gaussianity can be divided into two
cases.  When the linear term dominates over the second order one in Eq.~\eqref{eq:S32},
it is convenient to define a non-linearity parameter for isocurvature 
fluctuations as in the case of
adiabatic ones. We denote the non-linearity parameter as  $f_{\rm
NL}^{\rm (iso)}$ and  define it as the same as the counterpart in the
adiabatic case:
\begin{eqnarray}
\label{eq:iso_linear}
S_{3/2} = S_g + f_{\rm NL}^{\rm (iso)} S_g^2,
\end{eqnarray}
where $S_g$ is the Gaussian part of isocurvature fluctuations.  With
this definition, $f_{\rm NL}^{\rm (iso)}$ is calculated as
\begin{eqnarray}
f_{\rm NL}^{\rm (iso)} = { \Gamma'' / \Gamma \over \left( \Gamma' / \Gamma \right)^2}
- {1 \over 2}~.
\end{eqnarray}
Notice that $f_{\rm NL}^{\rm (iso)}$ takes 
almost the same value as $f_{\rm NL}$ for adiabatic
fluctuation although the sign is different.  However, even if the size
of $f_{\rm NL}^{\rm (iso)}$ is comparable to that of the counterpart
for the adiabatic mode, the signal from the bispectrum depends on the
combination of $\alpha^2 f_{\rm NL}^{\rm (iso)}$ and $\alpha f_{\rm
  NL}^{\rm (iso)}$ for the bispectrum coming from the 3-point function
of $\langle SSS \rangle$ and $\langle SS\zeta \rangle$, i.e., purely
isocurvature and correlated parts, respectively.  Thus they are
suppressed by the fraction of isocurvature perturbations $\alpha$,
which is severely constrained by observations. Hence non-Gaussianity
from isocurvature fluctuations cannot be large compared to the
adiabatic one in this case.

However, when the second order term dominates over the first order
one in Eq.~(\ref{eq:S32}), the argument becomes different.  In this kind of case, the
isocurvature fluctuations can be simply written as $S_{3/2} = S_g^2$,
which is called the quadratic model in \cite{Hikage:2008sk}.  (The
model characterized by Eq.~\eqref{eq:iso_linear} is called linear
model in \cite{Hikage:2008sk}.) In the quadratic model, its power
spectrum is determined by the second order term, thus non-Gaussianity
in this model can be represented only with $\alpha_{3/2}$, which
characterizes the size of the power spectrum of isocurvature
fluctuations relative to the adiabatic ones and is given by
\begin{equation}
%\label{ }
\alpha_{3/2} \equiv \frac{P_{S_{3/2}}}{P_\zeta + P_{S_{3/2}}}. 
\end{equation}
This fraction should be small to fit to current data of CMB power
spectrum, thus $P_\zeta \gg P_{S_{3/2}}$ and using
Eq.~\eqref{eq:coruncor} and $R \ll 1$, we obtain the following relation,
\begin{equation}
%\label{ }
\alpha_{3/2} = \frac{P_{S_{3/2}}}{(1+R)P_\phi + P_{S_{3/2}}}
\simeq \frac{1}{1+R} \frac{P_{S_{3/2}}}{P_\phi} \simeq \alpha_{\rm uncorr}.
\end{equation}
As discussed in the previous section, this parameter is constrained as
$\alpha_{\rm uncorr} < 0.072$ at 95 \% C.L. Thus we use this value as
a representative one in the following.

Here it should be noted that, even if we characterize the primordial
non-Gaussianity for both adiabatic and isocurvature fluctuations with
the nonlinearity parameters $f_{\rm NL}$ and $f_{\rm NL}^{\rm (iso)}$,
their evolutions  to the present epoch, which are
encoded in the transfer functions, are different. Thus the comparison
of the non-linearity between isocurvature and adiabatic
fluctuations is not so simple. However, some useful relations are
obtained among $f_{\rm NL}, f_{\rm NL}^{\rm (iso)}$ and $\alpha_{3/2}$
in \cite{Hikage:2008sk}.  Thus we make use of those relations.

Since non-Gaussianity from isocurvature fluctuations tends to be small
for the case with the linear model, as discussed above, we here consider
the quadratic case.  Non-Gaussianity from the uncorrelated term can be
estimated effectively as
\begin{equation}
\label{eq:iso_fNL}
f_{\rm NL} \simeq 30 \left( \frac{\alpha_{3/2}}{0.072} \right)^{1/2}, 
\end{equation}
where $f_{\rm NL}$ in the left hand side is the non-linearity parameter
for adiabatic fluctuations. This relation has been obtained by finding
$\alpha_{3/2}$ which gives the same S/N for adiabatic $f_{\rm NL}$
assuming WMAP5 noise. As seen from the above relation,
relatively large non-Gaussian fluctuations can be generated from
isocurvature fluctuations even if we take the value allowed by current
severe constraint on the isocurvature fluctuations.

Next we discuss non-Gaussianity from isocurvature fluctuations in the
curvaton scenario.  For the curvaton scenario, we considered two cases
for the generation of CDM and discussed isocurvature fluctuations for
each separetely. When CDM is created from the inflaton, isocurvature fluctuations
are given by Eq.~\eqref{eq:curvaton_iso_inf}, from which we can
evaluate the non-linearity parameter $f_{\rm NL}^{\rm (iso)}$ as
\begin{equation}
%\label{ }
f_{\rm NL}^{\rm (iso)} 
= - \left( \frac{1}{4f_{\rm dec}} - \frac{1}{3} - \frac{f_{\rm dec}}{6} \right) 
= -\frac{1}{5} f_{\rm NL}.
\end{equation}
Notice that $f_{\rm NL}^{\rm (iso)}$ is almost the same in size as the
adiabatic counterpart $f_{\rm NL}$ except from the sign, which is
similar to the case of the modulated reheating scenario.  
Although the size of the non-linearity parameters are almost
the same between $f_{\rm NL}$ and $f_{\rm NL}^{\rm (iso)}$, the signal
of the bispectrum is suppressed by the isocurvature fraction as
discussed above. Thus non-Gaussianity from the isocurvature
fluctuations in this case would be also small as well.

The other case we considered is that CDM is produced from the decay of
the curvaton. In this case, the isocurvature fluctuations are written
as Eq.~\eqref{eq:curvaton_iso_curvaton}, then the nonlinearity
parameter can be given by
\begin{equation}
%\label{ }
f_{\rm NL}^{\rm (iso)} =  -\frac{1}{4(1-f_{\rm dec})} 
\left( 
1 + 2 f_{\rm dec} + \frac{2}{3} f_{\rm dec}^2
\right). 
\end{equation}
As seen from this expression, when $f_{\rm dec}$ is close to 1,
$f_{\rm NL}^{\rm (iso)}$ would be very large.  Thus in this case, even
if the bispectrum itself is suppressed by the fraction of isocurvature
fluctuations, its signal can be very large. However, notice that the
sign of $f_{\rm NL}^{\rm (iso)}$ is negative.  In fact, when $f_{\rm dec}$
is close to 1, $f_{\rm NL}$ for the adiabatic fluctuations is $f_{\rm
  NL} \sim \mathcal{O}(1)$. Thus, when $f_{\rm dec} \sim 1$,
non-linearity mainly comes from isocurvature fluctuations and its size
can be large but the sign is negative.  Since too large value of
$f_{\rm NL}^{\rm (iso)}$ would be disfavored by observations,
this scenario may contradict even with current observations.

When the quadratic term dominated over the first order term, the
argument is the same as the case with modulated reheating, in which the
fraction of isocurvature fluctuation $\alpha$ gives the size of
non-Gaussianity or bispectrum.
Hence, when the quadratic term dominates in $\zeta_{(\sigma)}$,
non-Gaussianity from isocurvature fluctuations can be 
large in the curvaton scenario as well.

%%%%%%%%%%%%%%%%%%%%%%%%%%%%%
\section{Summary and Discussion}\label{sec:sum}
%%%%%%%%%%%%%%%%%%%%%%%%%%%%%

In this paper, we considered DM isocurvature fluctuations and
non-Gaussianity in models where the adiabatic curvature fluctuations can
be produced not only from of a light scalar field other than
inflaton (modulus or curvaton), but from the inflaton fluctuations,
which is called a mixed scenario.  Regarding the non-Gaussianity of the
curvature (adiabatic) fluctuations, we have found that relatively large
non-Gaussianity can be realized as $f_{\rm NL} \sim \mathcal{O}(10-100)$
without conflicting with the current constraints on the fraction of the
CDM isocurvature fluctuations. In other words, the current limit on
isocurvature fluctuations is not severe enough to prohibit large $f_{\rm
NL}$ for such mixed scenarios. However, for the future CMB experiments
such as Planck satellite, the limit on the uncorrelated isocurvature
fluctuations will be improved as $\alpha_{\rm uncorr} \lesssim 0.01$
\cite{Enqvist:1999vt,planck:2006uk,Bucher:2000kb}. We showed that this
projected limit translates into the bound on the non-linearity parameter
as $f_{\rm NL} < 3$.  

In fact, although we have mainly discussed
non-Gaussianity in the curvature (adiabatic) fluctuations, nonlinearity
can also arise from isocurvature fluctuations.  We have also discussed
 non-Gaussianity of this type and found that it can be large as
shown in Eq.~\eqref{eq:iso_fNL}, which corresponds to the size of the
adiabatic nonlinearity parameter as $f_{\rm NL} \sim 10$ even with the
projected Planck limit for $\alpha_{\rm uncorr}$.  However, it should be
noted that the signature in the bispectrum of adiabatic and isocurvature
 fluctuations are not the same, thus we may differentiate non-Gaussianity from these
fluctuations to some extent\footnote{
The phase difference between adiabatic and isocurvature 
fluctuations in acoustic oscillations can help to distinguish non-Gaussinity 
from these two modes \cite{Hikage:2009rt}.
}.  Thus, if we find that the
value of the local type (adiabatic) non-linearity parameter is large in
the future experiments, cosmological scenarios with gravitino DM  
may be disfavored  even if we consider a mixed fluctuation scenario.

Furthermore, we have also investigated DM isocurvature fluctuations in
the framework of the curvaton, in particular focusing on a 
mixed scenario.  Our discussions for the curvaton case also
apply to a generic DM, which originates from the inflaton or
the curvaton. We have found that
large non-Gaussianity in the curvature (adiabatic) perturbations is
possible with the current level of isocurvature constraints in the curvaton case as well. However, as
in the case of the modulated reheating scenario, if the limit becomes
severer and large (adiabatic) non-Gaussianity of the local type is found in the future, 
DM are unlikely to be produced from the decay of the
inflaton or the curvaton in the curvaton scenario, which would give
important implications to the generation mechanism of DM.

%%%%%%%%%%%%%%%%%%%%%%%%%%%%%%%%%%%%%%%%%%
\subsection*{Acknowledgments}
%%%%%%%%%%%%%%%%%%%%%%%%%%%%%%%%%%%%%%%%%%

We thank Jun'ichi Yokoyama and Takahiro Tanaka for the collaboration at
the early stage. We also grateful to Masahiro Kawasaki, Kazunori Kohri,
and Fuminobu Takahashi for useful discussions. This work is supported by
JSPS Grant-in-Aid for Scientific research, No.\,19740145 (T.T.) and
No.\,21740187 (M.Y.). S.Y. is supported in part by Grant-in-Aid for
Scientific Research on Priority Areas No. 467 ``Probing the Dark Energy
through an Extremely Wide and Deep Survey with Subaru Telescope''.  He
also acknowledges the support from the Grand-in-Aid for the Global COE
Program ``Quest for Fundamental Principles in the Universe: from
Particles to the Solar System and the Cosmos '' from MEXT of Japan. We
would like to thank the organizers of the IPMU workshop on ``Focus week
on non-Gaussianities in the sky'' and the GCOE/YITP workshop
YITP-W-09-01 on ``Non-linear cosmological perturbations'' for their
hospitality, during which a part of this work was done.

\end{document}